# Enhanced Radiation Hardness of InAs/GaAs Quantum Dot Lasers for Space Communication

*Manyang Li[1,2,#], Jianan Duan[3,#], Zhiyong Jin[3,#], Shujie Pan[1,4,#], Wenkang Zhan[1,2], Jinpeng Chen[5], Jinling Yu[5], Xiaotian Cheng[6], Zhibo Ni[6], Chaoyuan Jin[6], Tien Khee Ng[7], Jinxia Kong[1,2], Xiaochuan Xu[3], Yong Yao[3], Bo Xu[1,2], Siming Chen[1,2], Zhanguo Wang[1,2], Chao Zhao[1,2,*]*

[1] Laboratory of Solid State Optoelectronics Information Technology, Institute of Semiconductors, Chinese Academy of Sciences, Beijing 100083, China

[2] College of Materials Science and Opto-Electronic Technology, University of Chinese Academy of Sciences, Beijing 101804, China

[3] National Key Laboratory of Laser Spatial Information, Guangdong Provincial Key Laboratory of Integrated Photonic-Electronic Chip, School of Integrated Circuits, Harbin Institute of Technology, Shenzhen 518055, China

[4] HS Photonics Co., Ltd., Xiangjiang Science & Technology Innovation Base, Changsha, Hunan 413000, China

[5] Institute of Micro/Nano Devices and Solar Cells, School of Physics and Information Engineering, Fuzhou University, Fuzhou 350108, China

[6] College of Information Science and Electronic Engineering, State Key Laboratory of Silicon and Advanced Semiconductor Materials, Zhejiang University, Hangzhou, China

[7] Photonics Laboratory, King Abdullah University of Science and Technology (KAUST), Thuwal 23955-6900, Saudi Arabia

*Email: zhaochao@semi.ac.cn

#Equally contributing authors

## Abstract

Semiconductor lasers have great potential for space laser communication. However, excessive radiation in space can cause laser failure. In principle, quantum dot (QD) lasers are more radiation-resistant than traditional semiconductor lasers because of their superior carrier confinement and smaller active regions. However, the multifaceted nature of radiation effects on QDs resulted in ongoing controversies. Comprehensive testing under simulated space conditions is also necessary to validate their performance. In this work, we conduct radiation tests on various In(Ga)As/GaAs QD and quantum well (QW) materials and devices. Our results reveal that InAs/GaAs QDs with filling factors greater than 50% exhibit greater radiation hardness than those below 50%. Furthermore, most InAs/GaAs QDs show superior radiation resistance compared to InGaAs/GaAs QW when exposed to low proton fluences of $1 \times 10^{11}$ and $1 \times 10^{12}$ $cm^{-2}$, resulting from radiation-induced defects. The linewidth enhancement factor (LEF) of well-designed QD lasers remains remarkably stable and close to zero, even under proton irradiation at a maximum fluence of $7\times10^{13}$ $cm^{-2}$, owing to their inherent insensitivity to irradiation-induced defects. These QD lasers demonstrate an exceptional average relative intensity noise (RIN) level of -162 dB/Hz, with only a 1 dB/Hz increase in RIN observed at the highest fluence, indicating outstanding stability. Furthermore, the lasers exhibit remarkable robustness against optical feedback, sustaining stable performance even under a feedback strength as high as -3.1 dB. These results highlight the significant potential of QD lasers for space laser communication applications, where high reliability and resilience to radiation and environmental perturbations are critical.

## Introduction

Space laser communication has emerged as a game-changer in the field, offering numerous advantages over traditional microwave communication. Its larger capacity and enhanced security make it a vital tool for application in space communication.[1,2] Laser communication, particularly among small satellites, has transformed satellite network coverage and significantly reduced transmission delays, enabling the development of independent satellite networks. Many countries have confirmed the feasibility of laser communications between various orbits and the ground,[3] with successful tests demonstrating the transmission of over a hundred gigabytes of data.[4] Semiconductor lasers play a critical role as light sources or pumping lasers in space laser communication systems due to their high efficiency, long lifetime, and ease of modulation.[5] However, high-energy particles like protons, electrons, and heavy ions in the space environment adversely affect the performance of these minority carrier devices, which are particularly sensitive to radiation-induced defects than other components in the communication systems.[6] It is crucial to study the radiation hardness of semiconductor lasers to overcome the challenges posed by the space radiation environment.

Semiconductor quantum dot (QD) lasers have unique properties that make them promising candidates for space laser communication. These lasers exhibit higher temperature stability and operate with lower threshold currents, which helps to minimize power requirements for missions in space where the temperatures can vary dramatically, ranging from -120 °C to 120 °C.[7] Their high insensitivity to reflections eliminates the need for isolators, further enhancing the compactness of space laser communication systems.[8] In theory, the smaller active area and stronger carrier confinement in QDs reduce radiation-induced carrier migration to nonradiative recombination centers, compared to bulk material and quantum well (QW) lasers. This characteristic can ultimately improve the radiation hardness of QD lasers.[9-13]

However, reports indicate a decrease in photoluminescence (PL) intensity in QDs after irradiation,[14-17] while some studies show an increase in PL intensity for InAs/GaAs QDs when irradiated with a specific energy or fluence, highlighting the complexity of this phenomenon.[18,19] The role of radiation in defect formation within QDs is still debated. O'Driscoll et al. suggested that displacements occur within the dots, leading to the aggregation of these displacements into a single defect.[20] In contrast, A. Cavaco et al. found no evidence of defect generation in QDs at a proton fluence of $10^{14}$ $cm^{-2}$.[21] Controversy also exists regarding the impact of QD density on radiation hardness, with conflicting findings reported.[20,22] These variations highlight the multifaceted nature of radiation effects on QDs, shaped by material properties, structural parameters, and radiation conditions. There is a need to address unresolved and long-standing controversies to fully leverage the potential of QD lasers for space-based applications. Comprehensive testing under simulated space conditions is also necessary to validate their performance.

This report aims to have a comprehensive understanding of the radiation hardness of different QDs and the lasers. We introduce the filling factor, which is the ratio of the area covered by QDs to the total area of the active layer, accounting for both QD density and QD size. Our PL characterization reveal that InAs/GaAs QDs with filling factors greater than 50% exhibit greater radiation hardness than those with filling factors below 50%. Furthermore, most InAs/GaAs QDs show superior radiation resistance compared to InGaAs/GaAs QW when exposed to low proton fluences of $1 \times 10^{11}$ and $1 \times 10^{12}$ $cm^{-2}$. The temperature-dependent PL (TDPL) analysis reveal that the active energy of all samples decreases at a maximum proton fluence of $7 \times 10^{13}$ $cm^{-2}$, suggesting that radiation induces defects. Dislocations are further observed near the QWs but not in the QDs by transmission electron microscopy (TEM). QD lasers exhibit reduced efficiency and an increased threshold current under high proton fluences. However, they still deliver sufficient output power to meet operational requirements. Remarkably, the LEF of these lasers remains stable and close to zero, even at a maximum proton fluence of $7\times10^{13}$ $cm^{-2}$, owing to their insensitivity to irradiation-induced defects. Furthermore, these lasers achieve an average RIN level of -162 dB/Hz. Even at the highest fluence, the RIN increases by only 1 dB/Hz, demonstrating exceptional stability. Notably, their performance remains robust even under strong optical feedback, with a feedback strength as high as -3.1 dB. These characteristics make well-designed QD lasers excellent candidates for space laser communication applications, where reliability and resilience are paramount.

## Results

The QD samples are numbered from QD 1 to QD 5, while the QW sample for comparison is designated as QW 6. Sample QD 1 to 4 consist of single-layer QD, whereas QD 5 contains five layers. It is known that both QD density and size have a significantly impact on the properties of the materials. However, controlling one variable while varying another can be challenging. To address this, the filling factor was utilized to account for both QD density and size. Atomic force microscopy (AFM) images of QD 1 to 5 are shown in Figure 1(a)~(e), and the specifications of the QD samples, as obtained from the AFM analysis, are provided in Figure 1(f). Each QD sample features different filling factors.

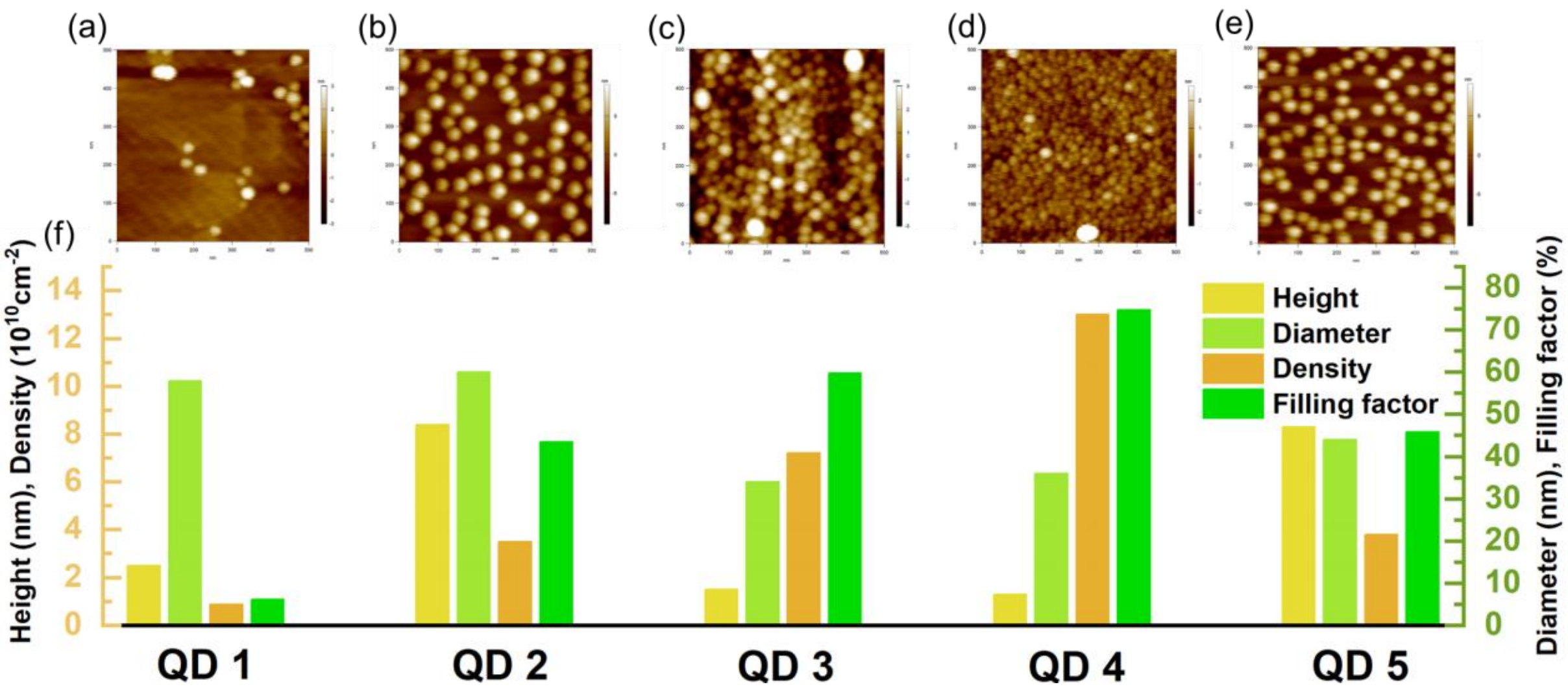

**Figure 1:** (a)~(e) The AFM images for QD 1 ~ 5. (f) The filling factors, densities, diameters, and heights of QDs.

Radiation primarily causes two types of damage to the crystal structure of semiconductors: displacement damage and ionization damage. Among these, displacement damage is the most significant degradation mechanism.[23] To study these effects, protons are used to investigate displacement damage, while $^{60}$Co γ-rays are employed to assess ionization damage hardness. All samples were exposed to $^{60}$Co γ-ray radiation doses ranging from 10 to 12000 Gy. The normalized integral PL intensity change for these samples remained consistently within 20% and no significant degradation occurred, indicating a good hardness to the total dose effect (see Supplementary Information for the result of the $^{60}$Co γ-rays radiation experiments, S1). Our research focuses specifically on displacement damage caused by proton radiation.

In our study, we have selected protons with an energy of 3 MeV to assess the radiation hardness of InAs/GaAs QDs and InAs/InGaAs QWs, which is common in most proton radiation experiments.[16,18,21,24,25] The setup for radiation experiment and the sample structures are shown in Figure 2(a). The radiation beam produced by proton source is accelerated by tandem electrostatic accelerator and directed perpendicular to the surface of the samples. Satellites typically operate in low Earth orbit, where the displacement damage dose is equivalent to 9.71 $\times 10^7$ to 3.87 $\times 10^9$ (10 MeV protons) • $cm^{-2}$ per year, assuming a 3 mm aluminium shield thickness.[26] When converted to 3 MeV protons, the fluence ranges from approximately 3.24 $\times 10^7$ to 1.29 $\times 10^9$ $cm^{-2}$ per year.[27] Given that the typical lifetime of satellites is between 7 to 10 years,[28] a proton fluence of $1 \times 10^{11}$ $cm^{-2}$ can be used to evaluate the radiation hardness of these structures in Low Earth orbit. Our study used proton fluence levels ranging from $1 \times 10^{11}$ to $7 \times 10^{13}$ $cm^{-2}$ to investigate displacement damage.

The impact of radiation on the PL spectra of all samples is significant. In Figure 2(b), we observe the typical evolution of PL spectra at various proton radiation fluences. As the fluence increases, the PL intensity decreases; however, the overall shape of the curve remains largely unchanged (see Supplementary Information for the PL spectra of QD 1~5 and QW 6 at $1 \times 10^{13}$ $cm^{-2}$ proton radiation fluences, S2). In Figure 2(c), we present the integrated PL intensities of QD 1~5 and QW 6 after exposure to proton radiation. To isolate the impact of radiation fluences on PL evolution, we first eliminate the influence of PL nonuniformity among pieces from the same sample. We achieve this by measuring the PL intensity of each sample before radiation, using the intensity of the sample without radiation as the standard. The ratio of the PL intensity of other samples before radiation to that of the zero-radiation sample reflects the unique characteristics of each sample. The PL intensity measured after radiation for each sample is then divided by this ratio, ensuring that the differences in PL intensity after radiation accurately represent the effects of varying radiation fluences. As shown in Figure 2(c), QD samples with filling factors greater than 50% (QD 3 and 4) exhibit better radiation tolerance. At low proton fluences of $1 \times 10^{11}$ $cm^{-2}$ and $1 \times 10^{12}$ $cm^{-2}$, most QD samples exhibit superior radiation hardness compared to QW samples. However, at a proton fluence of $1 \times 10^{13}$ $cm^{-2}$, the QD 1 and 2 show worse radiation hardness compared to the QW samples. In contrast, QD 3, 4, and 5 show similar radiation hardness to the QW samples. Moreover, at high proton fluences of $4 \times 10^{13}$ and $7 \times 10^{13}$ $cm^{-2}$, the integrated PL intensities of all samples decline.

To further confirm the reason of the PL degradation, QD and QW samples were analyzed before and after exposure to $1 \times 10^{11}$ $cm^{-2}$ proton radiation using TEM to compare the creation of defects near or within QDs and QW. Figure 2(d) shows that the lattice mismatch causes significant strain, distorting the lattice structure over a large area that extends several nanometers into the GaAs layer both above and below the QDs.[29] Figure 2(e) shows the shape of the QDs changes after radiation, likely due to compressive strain from radiation-created defects.[14] As shown in Figure 2(f), the QW layer appears wavy for the sample before radiation, indicating inhomogeneous strain.[30] After radiation exposure, dislocations form near the upper interface of the InGaAs QW and GaAs, creating sharp corners, as shown in Figure 2(g). These dislocations act as nonradiative sinks for excess charge carriers, potentially contributing to the 10.5% decrease in integrated PL intensity shown in Figure 2(c). Unlike in the QWs, radiation does not induce dislocations near or within the QDs. It indicates a higher kinetic energy barrier for dislocation formation in QDs, supporting the conclusion that QDs demonstrate better radiation hardness.

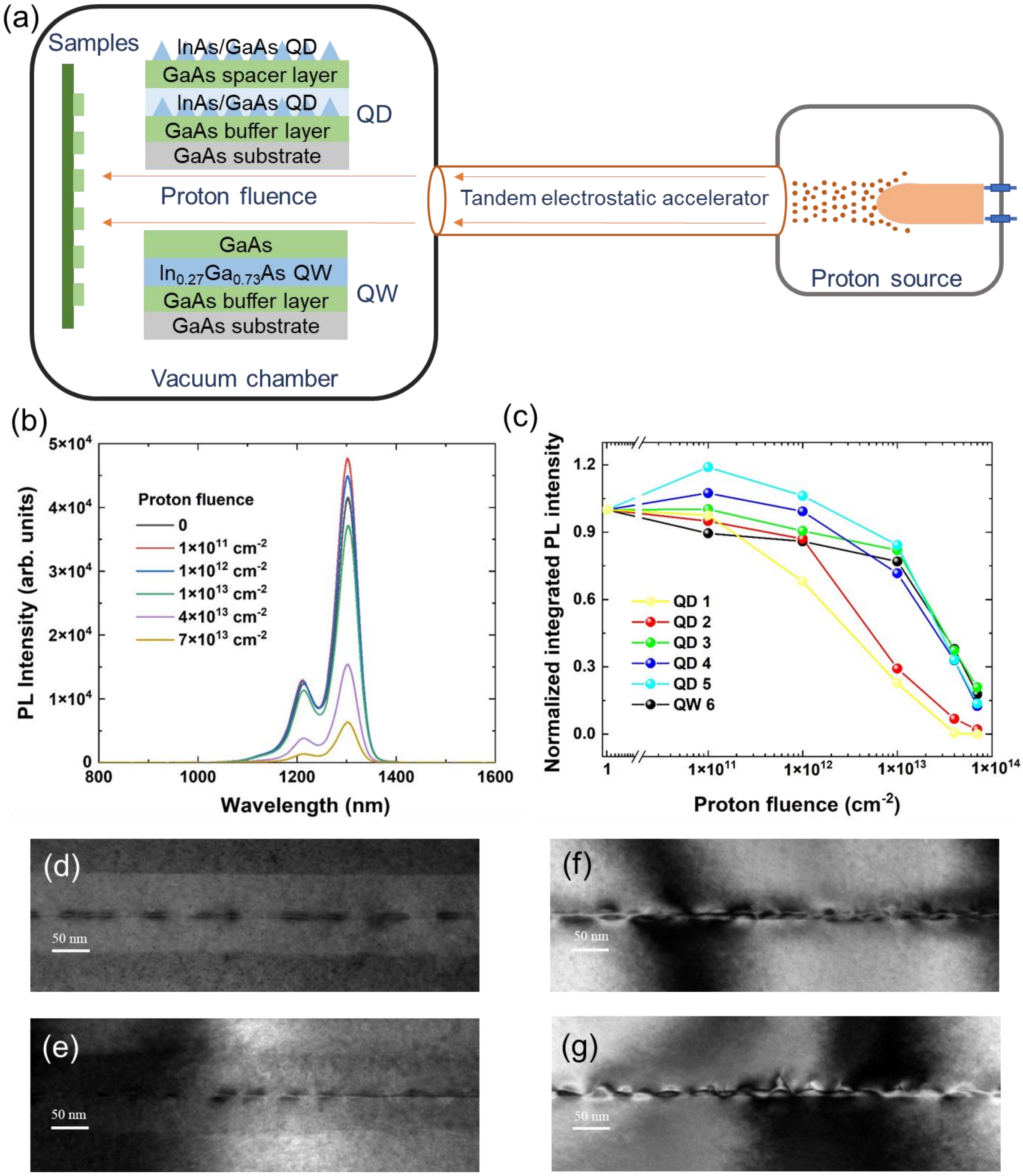

**Figure 2** (a) The setup for radiation experiment and the sample structures of QD 1~5 and QW 6. (b) Typical PL spectra at different proton fluence. (c) Normalized integrated PL intensities of QD 1~5 and QW 6 as a function of proton fluence. The TEM characterization of (d) QD 2 without radiation. (e) QD 2 with proton radiation fluence of $1\times10^{11}$ cm$^{-2}$. (f) QW 6 without radiation. (g) QW 6 with proton radiation fluence of $1\times10^{11}$ cm$^{-2}$.

Figures 3(a) to (c) show the schematics of QW and QD samples with filling factors below and above 50%, respectively. QD samples with lower filling factors typically exhibit larger QD sizes and fewer QDs than those with higher filling factors, as evidenced by the AFM images in Figure 1. As shown in Figures 3(d) to (f), at low radiation levels, carriers in QWs can easily migrate to radiation-induced defects, resulting in nonradiative recombination and a reduction in PL intensity. In contrast, the decrease in PL intensity for QD samples at low radiation levels is mainly due to carriers escaping from QDs into the wetting layers for nonradiative recombination, since there are relatively few defects within the QDs. As a result, QDs exhibit improved radiation hardness at low radiation levels than QWs, thanks to their carrier confinement capabilities.[25,31,32] In samples with high filling factors, the ratio of QDs containing defects is lower than in those with low filling factors. This is attributed to their smaller size and larger number of QDs, which together help minimize the effects of nonradiative recombination and improve their radiation hardness. Therefore, at low radiation levels, QD samples with high filling factors exhibit better radiation hardness than those with low filling factors, and both outperform QW samples. Figures 3(g) to (i) show that at a high radiation fluence of $1 \times 10^{12}$ cm$^{-2}$, the density of defects induced by radiation is significantly high. Even when comparing QD samples with high filling factors to QW samples, QDs maintain their carrier confinement, preventing carriers from migrating to defects. As a result, the QD sample with a high filling factor demonstrates superior radiation tolerance compared to the QW sample. However, this advantage diminishes as the radiation fluence increases. At higher radiation fluences, ranging from $1 \times 10^{13}$ cm$^{-2}$ to $7 \times 10^{13}$ cm$^{-2}$, the PL intensities of QD samples with high filling factors decline similarly to those of QW samples. This decline is due to the high density of defects present in both QDs and the wetting layer, which allows for nonradiative carrier recombination at these defects despite the initial advantages of carrier confinement in the QDs.

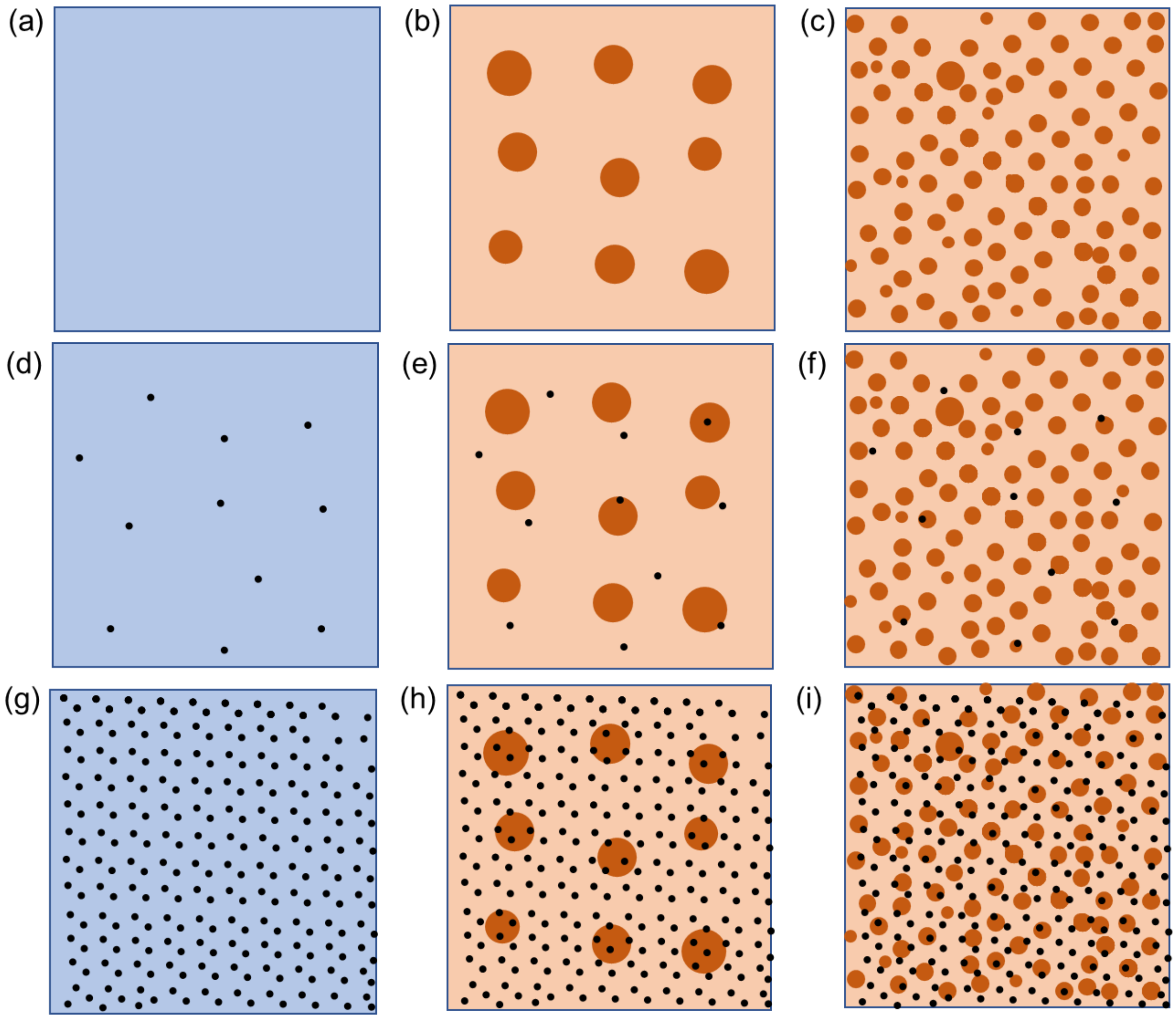

**Figure 3** Schematic diagrams of the active regions, with (a) ~ (c) no radiation, (d) ~ (f) low radiation fluence, and (g) ~ (i) high radiation fluence, of QW samples (blue color, 1st column), QD samples with low filling factors (2nd column), and QD samples with high filling factors (3rd column), respectively. The black dots represent defects created by radiation, and the brown dots represent the QDs.

As a result, the PL intensities of QD samples with low filling factors decrease more quickly as the radiation fluences increase, as shown in Figure 2(b). As radiation fluences increase, the potential for QDs to contain defects rises. Therefore, the carrier confinement of QDs may deteriorate PL. The QD samples with low filling factors typically have fewer and larger QDs, which results in a higher likelihood of defects. As a result, for radiation fluences exceeding 1 × $10^{12}$ $cm^{-2}$, the radiation hardness of QD samples with low filling factors (QD 1 and 2) is notably lower than that of the QW 6. Moreover, QD 5, with five layers of QDs in the active region, shows greater radiation tolerance than the material with only one layer of QDs at every fluence tested (QD 2), despite both having similar QD densities and filling factors. This enhanced tolerance can be attributed to improved carrier confinement in the five-layer QD structure, which helps prevent carriers from escaping to the GaAs layer and undergoing nonradiative recombination. Additionally, the five-layer QD structure possesses a higher QD volume density than the single-layer one, further reducing nonradiative recombination outside the QD. The intense strain fields in the five-layer QD structure also contribute to this increased confinement.[33] These findings are consistent with observations that QDs embedded in a superlattice show improved hardness to proton radiation.[16] Interestingly, the PL enhancement observed after radiation at a relatively low fluence in QD 4 and 5 can be explained by the defects created by radiation, which enhance the efficiency of carriers transfer to the QDs.[10,25]

We also performed temperature-dependent PL measurements on all samples, both before and after exposing them to proton radiation, to evaluate any changes in the activation energy ($E_a$), which could indicate the creation or elimination of defects. Figure 4(a) shows the temperature-dependent PL spectra for QD 2 without radiation. The PL intensity remains relatively constant at low temperatures but decreases as the temperature increases (see Supplementary Information for the temperature-dependent PL spectra for QD 1~5 and QW 6 without radiation, S3). An Arrhenius plot of the integrated PL intensity for QD 2 before and after radiated with protons is shown in Figure 4(b). The integrated PL intensity for each sample remains relatively consistent or slightly fluctuates between 10 and 100 K, gradually decreasing and then exhibiting an exponential decrease after 100 K. This decrease is attributed to the thermal escape of photogenerated carriers from QDs to the wetting layer or the barrier for the QD samples, and from QWs to the GaAs barrier for the QW sample (see Supplementary Information for the Arrhenius plot of the integrated intensity of temperature-dependent PL for QD 1~5 and QW 6, S4).[34,35]

The $E_a$ can be determined from the Arrhenius plot of the integral intensity of temperature-dependent PL at high temperatures. As shown in Figure 4(c), $E_a$ generally decreases with increasing proton fluence, indicating that the creation of defects in the sample after radiation reduces carrier confinement. The $E_a$ of QW samples does not decrease more than that of QD samples after radiation; this may be because the $E_a$ of QW samples is already relatively low compared to that of QD samples before radiation. At a proton fluence of $1\times10^{11}$ $cm^{-2}$, the most significant decrease in $E_a$ is observed in QD 1, which shows a reduction of 52%. Furthermore, QD 5 and QW 6 show an increase in $E_a$, possibly due to the combined effects of annihilation and creation of defects caused by radiation. At a proton fluence of $7\times10^{13}$ $cm^{-2}$, all samples show a reduction of $E_a$. Moreover, the exponential decrease in integrated PL intensity shifts towards lower temperatures with higher proton fluence, as depicted in Figure 4(b) and Figure S4 of the Supplementary Information, since the creation of defects by radiation lowers the temperature required for carriers to reach the wetting layer or the barrier in the QD samples, or the GaAs barrier in the QW sample, increasing the likelihood of non-radiative recombination. This trend indicates that numerous defects have been created in the materials.[14]

To compare the radiation hardness of QD and QW, it is crucial to examine how radiation impacts carrier lifetime ($\tau$). We conducted TRPL characterization on QD 2 and QD 3, which have a dot density in the range of that of mature QD lasers, as well as QW 6. This was performed before and after exposure to a proton radiation dose of $1 \times 10^{11}$ $cm^{-2}$ to observe its influence on $\tau$.[36] QD 3 was explicitly included to evaluate the effect of radiation on $\tau$ for QDs with varying filling factors. The TRPL spectra and single exponential fittings for QD 2, 3, and QW 6, both before and after exposure to proton radiation, are shown in Figures 4 (d). After proton radiation, both QD and QW samples exhibited nearly identical τ values, consistent with other reports.[22] As shown in Figure 4(e), there was a slight decrease in τ for QD 2, 3, and QW 6 following proton radiation, likely due to non-radiative recombination resulting from radiation-induced defects. The less degradation of $\tau$ in QD samples compared to QW samples can be explained by the smaller active area and stronger carrier confinement in QDs. For QD 2, the lower degradation percentage of $\tau$ compared to the integrated PL intensity shown in Figure 2(c) may be due to nonradiative recombination occurring at radiation-induced defects in the wetting layer, reducing the capture of carriers into the dots and consequently lowering the carrier population in the dots. In QD 3, the lower degradation percentage of the integrated PL intensity shown in Figure 2(c) compared to $\tau$ may be due to that the defects enhance the efficiency of carriers transfer to the QDs thus increasing the carrier population in the dots.[10,25] In QW 6, the reduced degradation of $\tau$ compared to the integrated PL intensity shown in Figure 2(c) may result from nonradiative recombination induced by defects in the GaAs layer, which also diminishes the carrier density in the QW.

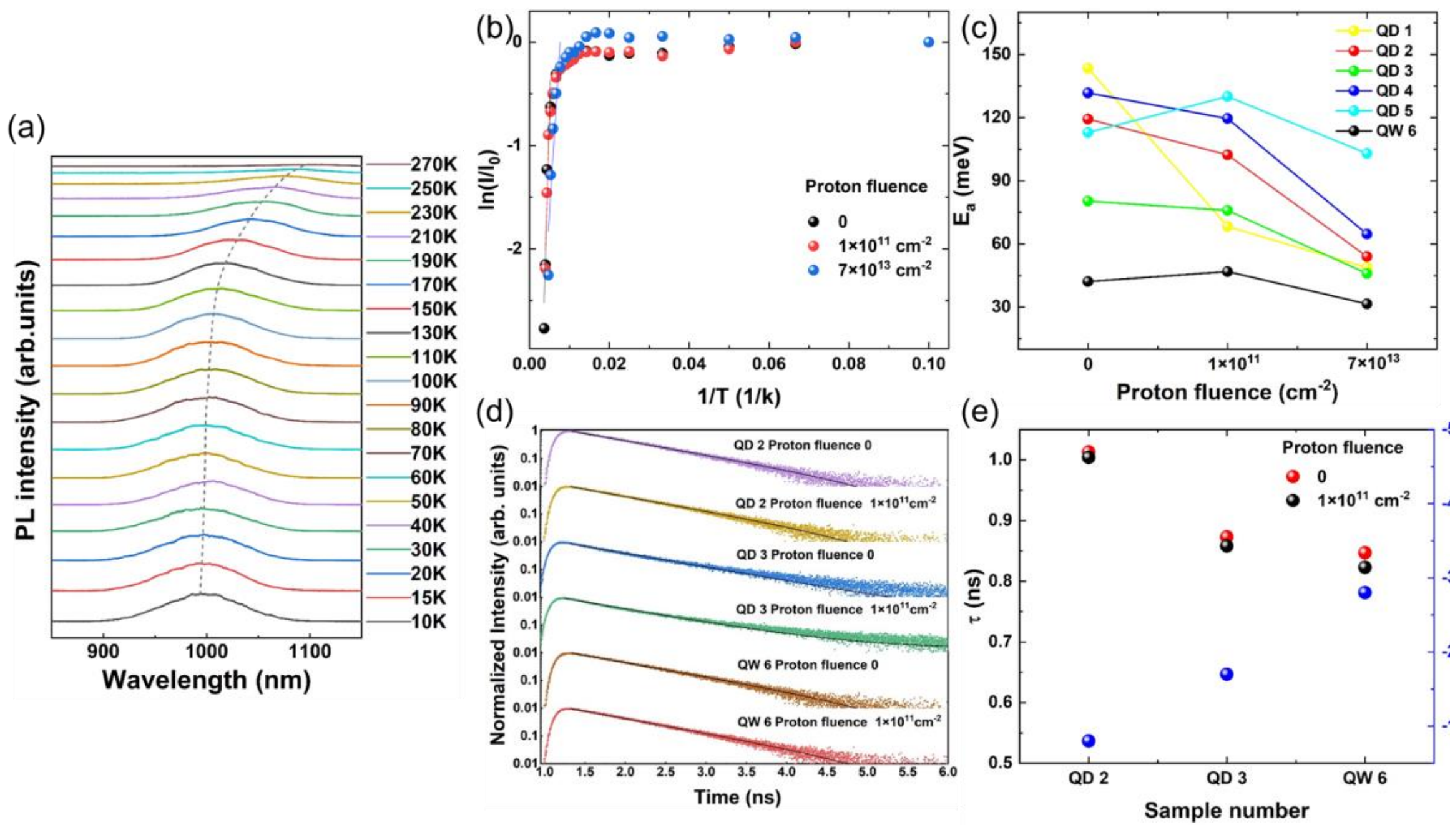

**Figure 4:** (a) The temperature dependent PL spectra for QD 2 before-radiation from 10 to 270 K. (b) Arrhenius plots of the integrated PL intensity of temperature-dependent PL carried out

on the QD 2 before-radiated, radiated with protons of fluences $1\times10^{11}$ $cm^{-2}$ and $7\times10^{13}$ $cm^{-2}$ ($I_0$ is the integrated PL intensity recorded at 10 K). (c) The active energies calculated from Arrhenius plots. (d) The time-resolved PL spectra and single exponential fitting of QD 2, 3 and QW 6 without and with proton radiation fluence of $1\times10^{11}$ $cm^{-2}$. (e) The carrier lifetime for QD 2, 3 and QW 6 without and with proton radiation fluence of $1\times10^{11}$ $cm^{-2}$.

The QD sample with the highest radiation hardness was selected for the growth and fabrication of QD lasers to evaluate their suitability for space laser communication applications. The RIN, defined as the ratio of the variance of laser intensity fluctuations to the average output power, is critical for determining the optical noise performance of laser sources. RIN primarily arises from intrinsic optical phase and frequency fluctuations caused by spontaneous emission and carrier noise. High RIN levels in laser sources can significantly degrade transmission performance in space communication systems by lowering the signal-to-noise ratio (SNR) and increasing the bit error rate (BER). In space laser communication, additional challenges arise from the interaction of laser sources with external optical feedback, which can destabilize the lasers and induce complex nonlinear behaviors, such as periodic or chaotic oscillations. Unlike terrestrial systems, space environments impose strict constraints on the size, weight, and power consumption of communication systems, making conventional optical isolators impractical due to their significant insertion losses and fabrication complexities. Therefore, it is essential to develop laser sources that are inherently resilient to external optical feedback to ensure the stability and reliability of space communication systems. Given these considerations, the optical noise and feedback characteristics of the QD lasers were rigorously tested to assess their potential for space laser communication. These investigations are vital to ensure that the lasers can maintain stable operation and deliver high performance under the challenging environmental conditions of space. The results provide valuable insights into the capability of QD lasers to serve as robust, low-noise light sources for future space-based photonic systems.

Figure 5 (a) shows the experimental setup used to measure RIN and optical feedback sensitivity (see Materials and methods). Before evaluating the RIN and optical feedback sensitivity of the QD lasers, we compared their static performance before and after radiation exposure. Figure 5(b) shows the power-current (PI) curves for five QD lasers fabricated on the same wafer using identical processes and exposed to different levels of proton fluence. It should be noted that a QD laser was reserved as a reference to ensure that the changes in laser performance were solely caused by radiation, thereby eliminating any potential errors introduced by the experimental setup. The results show that higher proton fluence leads to a decrease in the slope of PI curve, which indicates reduced efficiency. Additionally, the threshold current increases with increasing proton fluence. Figure 5(c) illustrates the variation in threshold current for the five QD lasers before and after exposure to different proton fluence levels. All five lasers exhibit a consistent threshold current of approximately 12 mA prior to radiation. When exposed to proton fluence levels below $1\times10^{12}$ $cm^{-2}$, the threshold current remains largely unaffected by radiation. However, as the proton fluence increases to higher levels, a noticeable rise in threshold current was observed after radiation. Nevertheless, even at the maximum fluence of $7\times10^{13}$ $cm^{-2}$, the threshold current increases only from 12 mA to 23 mA, while the lasers continue to maintain sufficient output power. Figure 5(d) compares the LEF of the five QD lasers before radiation and after exposure to varying proton fluence levels. The LEF characterizes the relationship between variations in the refractive index and optical gain relative to changes in carrier density. It measures the coupling between amplitude and phase fluctuations caused by current injection in semiconductor lasers. The LEF significantly impacts various aspects of laser performance, including optical noise, optical feedback response, modulation efficiency, and nonlinear dynamic behavior.[37] The LEF values shown in the Figure 5(d) were determined using the amplified spontaneous emission (ASE) method, specifically at the gain peak. Remarkably, the QD lasers exhibit a near-zero LEF, underscoring the high uniformity of the QDs, which contributes to their exceptional high gain performance. A particularly remarkable observation is that the LEF remains consistently stable and near zero, even under proton irradiation at the maximum fluence of $7\times10^{13}$ cm$^{-2}$. Details on the wavelength dependence of the LEF are provided in Supplementary Information S5. This high stability indicates that although proton radiation can induce defects in the QDs, their spatially discrete distribution effectively mitigates the negative impact of these defects on the laser's gain, thereby maintaining overall performance.

Figure 5(e) presents the averaged RIN values of five lasers measured before and after exposure to varying levels of proton fluence. The RIN spectra are recorded at different bias currents above the threshold current until they reach the saturation point for RIN value of the laser (see Supplementary Information S5). The RIN is relatively high at low frequencies due to the significant influence of current source noise, thermal noise, and mode partition noise. As the frequency and bias current increase, RIN gradually decreases and eventually reaches a saturation point. The averaged RIN value is determined as the mean of the saturated RIN within the 10 GHz to 20 GHz range. The QD lasers demonstrate excellent optical noise performance, achieving an average RIN as low as -162 dB/Hz. After proton radiation, the RIN values remain stable for fluence levels up to $1\times10^{12}$ cm$^{-2}$. Even at the highest fluence of $7\times10^{13}$ cm$^{-2}$, the RIN increases by only 1 dB/Hz, indicating the lasers' strong resilience to radiation-induced degradation. Figure 5(f) presents the optical spectrum mapping of the QD laser as a function of feedback strength at the highest proton fluence of $7\times10^{13}$ cm$^{-2}$. The optical spectrum changes under optical feedback before and after radiation can be found in Supplementary Information S5. The results reveal that the QD laser maintains exceptional stability against optical feedback, even at the maximum remarkable feedback strength of -3.1 dB. This indicates that the onset of coherence collapse, which typically occurs beyond a critical feedback strength, does not occur within the feedback strength range considered in this experiment. The only noticeable effect is a slight redshift in the modal wavelength. These findings highlight the exceptional stability of the QD laser, attributed to its low LEF—a key feature for applications in space laser communication, where optical feedback insensitivity is crucial for ensuring reliable and high-performance operation in challenging environments.

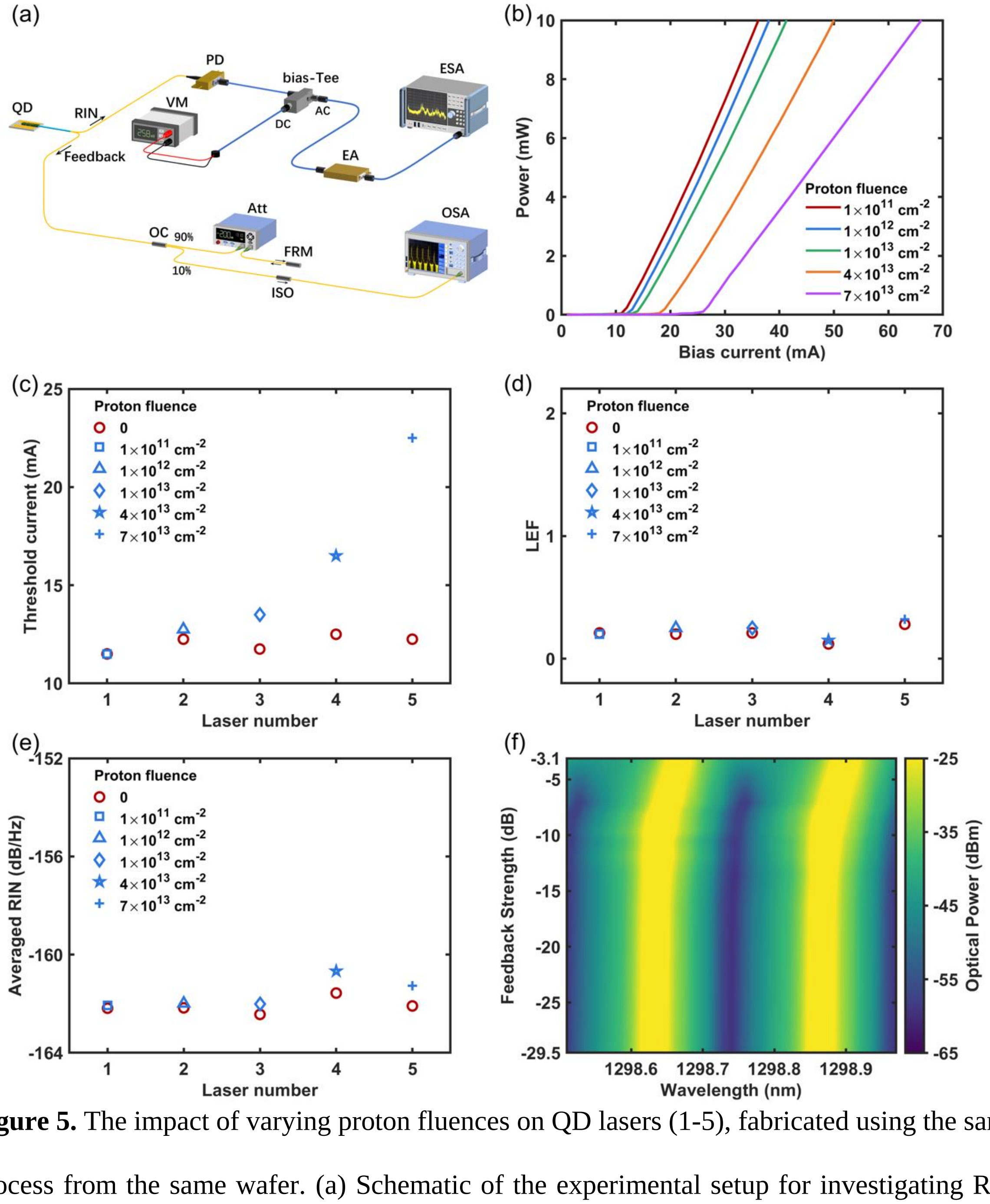

**Figure 5.** The impact of varying proton fluences on QD lasers (1-5), fabricated using the same process from the same wafer. (a) Schematic of the experimental setup for investigating RIN and optical feedback resistance. Abbreviations: QD, quantum dot laser; PD, photodiode; VM, voltmeter; Amp, electrical amplifier; ESA, electrical spectrum analyzer; Att, optical attenuator;

FRM, Faraday rotation mirror; ISO, isolator; OSA, optical spectrum analyzer. (b) The PI curves under different proton fluences, where each laser (1-5) was subjected to a specific fluence level. (c) Comparison of threshold currents before and after radiation, highlighting the effect of proton fluence on each laser (1-5). (d) Comparison of LEF before and after radiation for each laser (1–5) under varying proton fluences. (e) Averaged RIN values before and after radiation, illustrating the impact of proton fluence on each laser (1–5). (f) Optical spectra evolution under feedback strengths ranging from -29.5 dB to -3.1 dB, measured using laser 5 at a proton fluence of $7 \times 10^{13}$ $cm^{-2}$.

## Discussion

In conclusion, our findings demonstrate that QD structures exhibit greater radiation tolerance compared to QW structures. Specifically, QD samples with filling factors higher than 50% show improved radiation resistance over those with lower filling factors, and multiple layers of QDs provide better radiation tolerance. This is evidenced by the less degradation of PL intensity in QD samples with higher filling factors and with 5 layers. We identified that the proton fluence range from $1 \times 10^{11}$ $cm^{-2}$ to $1 \times 10^{12}$ $cm^{-2}$ allows QD samples to significantly outperform QW samples in terms of radiation tolerance. Most QD samples across various structures demonstrated less PL degradation than QW samples within this proton fluence range. Additionally, we observed defect creation in both QD and QW structures after radiation exposure, evidenced by a decrease in activation energy after $7 \times 10^{13}$ $cm^{-2}$ of proton radiation. The lasers show insenstitivty to radiation induce defects, evidenced by their LEF remains stable and close to zero. The QD lasers achieve an average RIN level of -162 dB/Hz and maintains exceptional stability, even at a remarkable feedback strength of -3.1 dB. QD lasers demonstrate significant potential as light sources for space communication systems, thanks to their remarkable radiation hardness, exceptional optical noise performance, and insensitivity to optical feedback, as highlighted in this paper. The use of QD lasers in space communications is set to revolutionize the way information is transmitted to and from space.

## Materials and methods

### Material growth

The samples were grown in a Riber 32P solid-source molecular beam epitaxy (MBE) system equipped with an arsenic (As) valve cracker, gallium (Ga), and indium (In) effusion cells. $As_4$ was used in the growth process, with the cracker temperature maintained at 600 °C. The substrate temperature was measured using a C-type thermocouple, and the growth temperature was calibrated by observing the transition of the reflection high-energy electron diffraction (RHEED) reconstruction pattern on the GaAs surface from (2 × 4) to c(4 × 4).[38] The GaAs growth rates were calibrated by monitoring the RHEED oscillations on GaAs substrates. Additionally, the growth rate of InAs QDs was calibrated by observing the transition from 2D to 3D growth mode through changes in the RHEED pattern along the [110] direction.[39]

Before transferring to the growth chamber, the samples underwent outgassing at 350°C in the buffer chamber. After the initial thermal cleaning, the oxide was removed from the n-GaAs (100) substrates by heating the sample to 620 °C in an $As_4$ atmosphere. A 200-nm GaAs buffer layer was firstly grown at 600 °C at a growth rate of 0.7 µm/h with an As beam equivalent pressure of $4.7 \times 10^{-6}$ Torr. Subsequently, the substrate temperature was reduced to grow InAs/GaAs low-dimensional structures, including a single-stacked InAs/GaAs QDs layer, a five-stacked InAs/GaAs QDs layer, and $In_{0.27}Ga_{0.73}As$/GaAs single quantum well structures, separately. To obtain QDs samples with varying densities, we implemented different growth conditions, including the substrate offset angle,[40] the InAs growth temperature, the InAs amount,[41] the InAs growth rate, and the $As_4$ flux.[42,43] A 200-nm GaAs layer was grown to attain a smooth surface. Finally, InAs/GaAs QDs were grown under the same conditions for AFM morphological and density analysis.

**Material characterization**

The morphologies of these structures were examined using AFM (MFP 3D Origin) after growth. The AFM images were processed using Asylum Research software to standardize the image contrast and measure the QD sizes. The QD densities for each sample were determined manually, and a Python program was used to calculate the filling factors.

Each material was cut into pieces measuring approximately 2 mm × 2 mm for exposure to different fluences of radiation. One piece was kept without exposure to radiation for comparison. The PL spectra of each piece at room temperature were analyzed using an iHR550 monochromator and an InGaAs detector cooled to -75 ℃. A solid-state laser with a wavelength of 532 nm was used for excitation, and the detection wavelength ranged from 800 nm to 1600 nm. Five spectra were collected for each piece to minimize errors from sample inhomogeneity. Before radiation, the room temperature PLs of all pieces were measured to identify any differences in PL intensities. These measurements were used as normalization factors for the PL spectra collected after the radiation experiments for comparison. The temperatures-dependent PL was measured 10 K to 270 K. The measurements were conducted using a helium cryostat (X20-OM) after samples were exposed to proton radiation at fluences of 0, $1 \times 10^{11}$, and $7 \times 10^{13}$ $cm^{-2}$. Three measurements were taken for each sample to minimize errors arising from sample non-uniformity.

The room-temperature PL and temperature-dependent PL spectra for the samples were analyzed using Origin to remove background signals and extract integrated intensities for the emission peaks. The background baselines were identified and averaged from spectral regions without emission, and then subtracted from the original spectra. After subtracting the baseline, these PL spectra are smoothed using the adjacent averaging method with 20 points.

After proton radiation, time-resolved PL was characterized for some samples at 77K after excitation by a short laser pulse of Chameleon Ti:sapphire laser from Coherent company (pulse duration 140 fs, repetition frequency 80MHz, central wavelength 720nm). The laser exciting power is 83 $\mu$W. The spectrometer is FHR 1000 from Horiba company. The PL is detected by ID230 Infrared Single-Photon detector. The multichannel picosecond event timer & TCSPC module is HydraHarp 400 from Picoquant company.

Some samples were examined using transmission electron microscopy (TEM) to characterize their cross-sections both before and after proton radiation. The sample preparation has been performed by a dual beam FIB microscope, Helios5CX (Thermofisher scientific). After selection of the target area, a protection layer of Pt was deposited on the exposed target surface inside the FIB microscope. Then the target area was FIB ablated (30 kV) into a prism-shaped (10 um length × 3 um width × 5 um depth). The sample was further mounted onto a Cu TEM grid and milled down to an 80 nm thin lamella. The TEM images were taken by a Talos F200s transmission electron microscope operated at 200 kV.

**Device fabrication**

The InAs/GaAs QD laser sample was grown and then made into a broad-area Fabry–Perot (FP) laser with a width of 50 μm and shallow ridge waveguides using standard photolithography and wet etching techniques. A 350-nm-thick layer of $SiO_2$ was deposited on the sample surface, and contact windows were opened on the ridge top. After that, Ti/Au was deposited on the top of the mesa. The substrate was then lapped down to a thickness of 125 μm, and n-type contacts (Ge/Au/Ni/Au) were deposited on the backside of the substrate. Finally, the laser bars were mounted on copper heatsinks and connected using gold wire bonding for testing.

**Relative intensity noise and optical feedback characterization**

The QD laser is driven by a direct current (DC) source, and its temperature is kept constant at 20 °C with the help of a thermo-electric cooler. In the RIN measurement configuration, the laser emission is coupled into a lensed fiber, where it is converted into an electrical signal by a low-noise photodiode with a 20 GHz bandwidth. The DC component of the photodiode output is recorded using a voltmeter, while the alternating current (AC) signal is amplified by a broadband amplifier with a typical gain of 30 dB before being analyzed with an electrical spectrum analyzer (ESA, Rohde&Schwarz FSW50).[44] For optical feedback sensitivity measurements, the laser emission is split by a 90/10 coupler , directing 90% of the power into the feedback path and reserving the remaining 10% for detection. In the feedback loop, a Faraday rotation mirror (FRM) is paired with a variable attenuator to generate controlled reflections. A polarization controller is used to compensate for fiber-induced dispersion and to ensure transverse electric (TE) polarization. The external cavity has a total length of 4 meters and consists of fiber components along with the required patch cables. The feedback strength is defined as the ratio of the returned laser power to the emitted laser power and can be adjusted using a variable attenuator to achieve a maximum value of -3.1 dB after accounting for coupling and connection losses. An isolator is positioned in the detection path to eliminate unwanted feedback, and the optical spectrum is analyzed using an optical spectrum analyzer (OSA, YOKOGAWA AQ6370D). This configuration ensures accurate and reliable measurements while minimizing signal degradation.[45]

## Radiation experiment

Proton radiation experiments were carried out using the EN-18 tandem electrostatic accelerator at Peking University. The protons had an energy of 3 MeV. The time and flux for each fluence for material samples are detailed in Tab 1. The fluences of proton radiation on lasers are the same as those on QD and QW samples, although the fluxes may vary depending on the accelerator’s condition.

**Table 1.** Time and flux for each proton fluence

| **Fluence ($cm^{-2}$)** | **$1 \times 10^{11}$** | **$1 \times 10^{12}$** | **$1 \times 10^{13}$** | **$4 \times 10^{13}$** | **$7 \times 10^{13}$** |
|---|---|---|---|---|---|

| Time (s) | 210 | 524 | 1118 | 4228 | 13782 |
|---|---|---|---|---|---|
| Flux ($cm^{-2}\cdot s^{-1}$) | $4.8 \times 10^8$ | $1.9 \times 10^9$ | $8.9 \times 10^9$ | $2.4 \times 10^9$ | $7.3 \times 10^8$ |

The γ-ray radiation experiment was carried out on the RSL2089 $^{60}Co$ source (produced by REVISS company) at Peking University. The dose rate and time for each dose are detailed in Tab 2.

**Table 2.** Dose rate and time for each dose

| Dose (Gy) | 10 | 100 | 1000 | 6000 | 12000 |
|---|---|---|---|---|---|
| Dose rate (Gy/s) | 0.5 | 0.5 | 0.5 | 0.5 | 0.5 |
| Time (s) | 20 | 200 | 2000 | 12000 | 24000 |

## Data availability

The data supporting the results reported in the article are provided with this paper.

## References

1 Svorec, R. W. & Gerardi, F. R. SPACE LASER COMMUNICATIONS - COMPETITION FOR MICROWAVES. *Photonics Spectra* **18**, 78-81 (1984).

2 Li, M. *et al.* Radiation hardness of semiconductor laser diodes for space communication. *Applied Physics Reviews* **11**, doi:10.1063/5.0188964 (2024).

3 Toyoshima, M. in *International Workshop "GOLCE2010"* (2010).

4 Xinhua. *China realizes ultra-high-speed intersatellite laser communications*, <https://global.chinadaily.com.cn/a/202401/12/WS65a11fb9a3105f21a507c0a8.html> (2024).

5 Evans, G., Leary, J. & Wilcox, J. APPLICATIONS OF SEMICONDUCTOR-LASERS IN SPACE COMMUNICATIONS. *Opt. Eng.* **22**, 247-255, doi:10.1117/12.7973092 (1983).

6 Boudenot, J.-C. in *Radiation Effects on Embedded Systems* (eds Raoul Velazco, Pascal Fouillat, & Ricardo Reis) 1-9 (Springer Netherlands, 2007).

7 Norman, J. C., Mirin, R. P. & Bowers, J. E. Quantum dot lasers—History and future prospects. *Journal of Vacuum Science & Technology A* **39**, doi:10.1116/6.0000768 (2021).

8 Lafon, R. *et al. Current status of NASA's Low-Cost Optical Terminal (LCOT) at Goddard Space Flight Center*. Vol. 12413 PWL (SPIE, 2023).

9 Aierken, A. *et al.* Optical properties of electron beam and -ray irradiated InGaAs/GaAs quantum well and quantum dot structures. *Radiation Physics and Chemistry* **83**, 42-47, doi:10.1016/j.radphyschem.2012.09.022 (2013).

10 Che, C. *et al.* Electron radiation effects on InAs/GaAs quantum dot lasers. *Laser Physics* **22**, 1317-1320, doi:10.1134/s1054660x12080051 (2012).

11 Gonda, S.-i. *et al.* Proton radiation effects in quantum dot lasers. *Applied Surface Science* **255**, 676-678, doi:10.1016/j.apsusc.2008.07.037 (2008).

12 Sobolev, N. A. Radiation effects in Si-Ge quantum size structure (Review). *Semiconductors* **47**, 217-227, doi:10.1134/S1063782613020188 (2013).

13 Sobolev, N. A. in *Handbook of Self Assembled Semiconductor Nanostructures for Novel Devices in Photonics and Electronics* (ed Mohamed Henini) 392-447 (Elsevier, 2008).

14 Sreekumar, R., Mandal, A., Chakrabarti, S. & Gupta, S. K. Effect of heavy ion implantation on self-assembled single layer InAs/GaAs quantum dots. *Journal of Physics D: Applied Physics* **43**, doi:10.1088/0022-3727/43/50/505302 (2010).

15 Piva, P. G. *et al.* Enhanced degradation resistance of quantum dot lasers to radiation damage. *Applied Physics Letters* **77**, 624-626, doi:10.1063/1.127065 (2000).

16 Huang, M. B., Zhu, J. & Oktyabrsky, S. Enhanced radiation hardness of photoluminescence from InAs quantum dots embedded in an AlAs/GaAs superlattice structure. *Nuclear Instruments and Methods in Physics Research Section B: Beam*

*Interactions with Materials and Atoms* **211**, 505-511, doi:10.1016/s0168-583x(03)01516-7 (2003).

17 Mares, J. W., Harben, J., Thompson, A. V., Schoenfeld, D. W. & Schoenfeld, W. V. Gamma Radiation Induced Degradation of Operating Quantum Dot Lasers. *IEEE Transactions on Nuclear Science* **55**, 763-768, doi:10.1109/tns.2008.918743 (2008).

18 Sreekumar, R., Mandal, A., Gupta, S. K. & Chakrabarti, S. Effect of high energy proton irradiation on InAs/GaAs quantum dots: Enhancement of photoluminescence efficiency (up to ∼7 times) with minimum spectral signature shift. *Materials Research Bulletin* **46**, 1786-1793, doi:10.1016/j.materresbull.2011.07.048 (2011).

19 Upadhyay, S. *et al.* Effects of high-energy proton implantation on the luminescence properties of InAs submonolayer quantum dots. *Journal of Luminescence* **171**, 27-32, doi:10.1016/j.jlumin.2015.11.007 (2016).

20 O'Driscoll, I., Blood, P., Smowton, P. M., Sobiesierski, A. & Gwilliam, R. Effect of proton bombardment on InAs dots and wetting layer in laser structures. *Applied Physics Letters* **100**, doi:10.1063/1.4730964 (2012).

21 Cavaco, A. *et al.* Carrier dynamics in particle-irradiated InGaAs/GaAs quantum dots. *physica status solidi (c)* **0**, 1177-1180, doi:10.1002/pssc.200303033 (2003).

22 Marcinkevicius, S. *et al.* Changes in carrier dynamics induced by proton irradiation in quantum dots. *Physica B-Condensed Matter* **314**, 203-206, doi:10.1016/s0921-4526(01)01361-8 (2002).

23 Raya-Armenta, J. M., Bazmohammadi, N., Vasquez, J. C. & Guerrero, J. M. A short review of radiation-induced degradation of III–V photovoltaic cells for space applications. *Solar Energy Materials and Solar Cells* **233**, doi:10.1016/j.solmat.2021.111379 (2021).

24 Ribbat, C. *et al.* Enhanced radiation hardness of quantum dot lasers to high energy proton irradiation. *Electronics Letters* **37**, doi:10.1049/el:20010118 (2001).

25 Leon, R. *et al.* Effects of proton irradiation on luminescence emission and carrier dynamics of self-assembled III-V quantum dots. *IEEE Transactions on Nuclear Science* **49**, 2844-2851, doi:10.1109/tns.2002.806018 (2002).

26 Zhang, H., Li, P. & Sun, Y. Investigation on Radiation Hardness Assurance for Components'Performance in Commercial Aviation Space. *Spacecraft Engineering* **28**, 81-86 (2019).

27 Gonda, S. i. *et al.* in *2007 IEEE 19th International Conference on Indium Phosphide & Related Materials.* 245-248.

28 Stewart, K. *Low Earth orbit*, <https://www.britannica.com/technology/low-Earth-orbit> (2024).

29 Vullum, P. E. *et al.* Quantitative strain analysis of InAs/GaAs quantum dot materials. *Scientific Reports* **7**, 45376, doi:10.1038/srep45376 (2017).

30 Yao, J. Y., Andersson, T. G. & Dunlop, G. L. Structure of lattice‐strained InxGa1−xAs/GaAs layers studied by transmission electron microscopy. *Applied Physics Letters* **53**, 1420-1422, doi:10.1063/1.99960 %J Applied Physics Letters (1988).

31 Leon, R. *et al.* Changes in luminescence emission induced by proton irradiation: InGaAs/GaAs quantum wells and quantum dots. *Applied Physics Letters* **76**, 2074-2076, doi:10.1063/1.126259 (2000).

32 Sobolev, N. A. *et al.* Enhanced radiation hardness of InAs/GaAs quantum dot structures. *Physica Status Solidi B-Basic Solid State Physics* **224**, 93-96, doi:10.1002/1521-3951(200103)224:1<93::Aid-pssb93>3.0.Co;2-6 (2001).

33 Ren, H.-W. *et al.* Confinement effects in strain-induced InGaAs/GaAs quantum dots. *Physica E: Low-dimensional Systems and Nanostructures* **7**, 403-407, doi:https://doi.org/10.1016/S1386-9477(99)00350-1 (2000).

34 Sreekumar, R., Mandal, A., Chakrabarti, S. & Gupta, S. H− ion implantation induced ten-fold increase of photoluminescence efficiency in single layer InAs/GaAs quantum dots. *Journal of Luminescence* **153**, 109-117 (2014).

35 Lambkin, J. D., Dunstan, D. J., Homewood, K. P., Howard, L. K. & Emeny, M. T. Thermal quenching of the photoluminescence of InGaAs/GaAs and InGaAs/AlGaAs strained‐layer quantum wells. *Applied Physics Letters* **57**, 1986-1988, doi:10.1063/1.103987 %J Applied Physics Letters (1990).

36 Nishi, K., Takemasa, K., Sugawara, M. & Arakawa, Y. J. I. J. o. S. T. i. Q. E. Development of Quantum Dot Lasers for Data-Com and Silicon Photonics Applications. *IEEE Journal of Selected Topics in Quantum Electronics* **23**, 1-7 (2017).

37 Grillot, F., Duan, J., Dong, B. & Huang, H. Uncovering recent progress in nanostructured light-emitters for information and communication technologies. *Light: Science & Applications* **10**, 156, doi:10.1038/s41377-021-00598-3 (2021).

38 Avery, A. R., Homes, D. M., Sudijono, J., Jones, T. S. & Joyce, B. A. The As-terminated reconstructions formed by GaAs(001): a scanning tunnelling microscopy study of the (2 × 4) and c(4 × 4) surfaces. *Surface Science* **323**, 91-101, doi:https://doi.org/10.1016/0039-6028(94)00635-0 (1995).

39 Tsukamoto, S., Honma, T., Bell, G. R., Ishii, A. & Arakawa, Y. Atomistic Insights for InAs Quantum Dot Formation on GaAs(001) using STM within a MBE Growth Chamber. *Small* **2**, 386-389, doi:https://doi.org/10.1002/smll.200500339 (2006).

40 Chen, M.-C., Harris Liao, M. C. & Lin, H.-H. Self-organized InAs/GaAs quantum dots grown on (100) misoriented substrates by molecular beam epitaxy. *Journal of Crystal Growth* **188**, 383-386, doi:https://doi.org/10.1016/S0022-0248(98)00059-1 (1998).

41 Joyce, P. B. *et al.* Optimizing the growth of 1.3 \ensuremath{\mu}m InAs/GaAs quantum dots. *Physical Review B* **64**, 235317, doi:10.1103/PhysRevB.64.235317 (2001).

42 Dubrovskii, V. G. *et al.* Dependence of structural and optical properties of QD arrays in an InAs/GaAs system on surface temperature and growth rate. *Semiconductors* **38**, 329-334, doi:10.1134/1.1682338 (2004).

43 Garcia, A. *et al.* Influence of As-4 flux on the growth kinetics, structure, and optical properties of InAs/GaAs quantum dots. *Journal of Applied Physics* **102**, 073526, doi:10.1063/1.2785969 (2007).

44 Wang, W. *et al.* Intensity noise reduction in quantum dot comb laser by lower external carrier fluctuations. *Opt Lett* **49**, 5007-5010, doi:10.1364/OL.532012 (2024).

45 Jin, Z. *et al.* Reflection sensitivity of dual-state quantum dot lasers. *Photon. Res.* **11**, 1713-1722, doi:10.1364/PRJ.494393 (2023).

## Acknowledgements

This work was supported by the National Key R&D Program of China (Grant No. 2023YFB2805900, S. M. C.; 2021YFB2206500, C. Z.), National Natural Science Foundation of China (Grant No. 62274159, C. Z.; 62204072, J. N. D), Beijing Natural Science Foundation (Grant No. L248103, C. Z.), CAS Project for Young Scientists in Basic Research (Grant No. YSBR-056, C. Z.), the “Strategic Priority Research Program” of the Chinese Academy of Sciences (Grant No. XDB43010102, C. Z.), and the Shenzhen Science and Technology Innovation Program (Grant No. JCYJ20240813104819027, J. N. D).

ToC figure

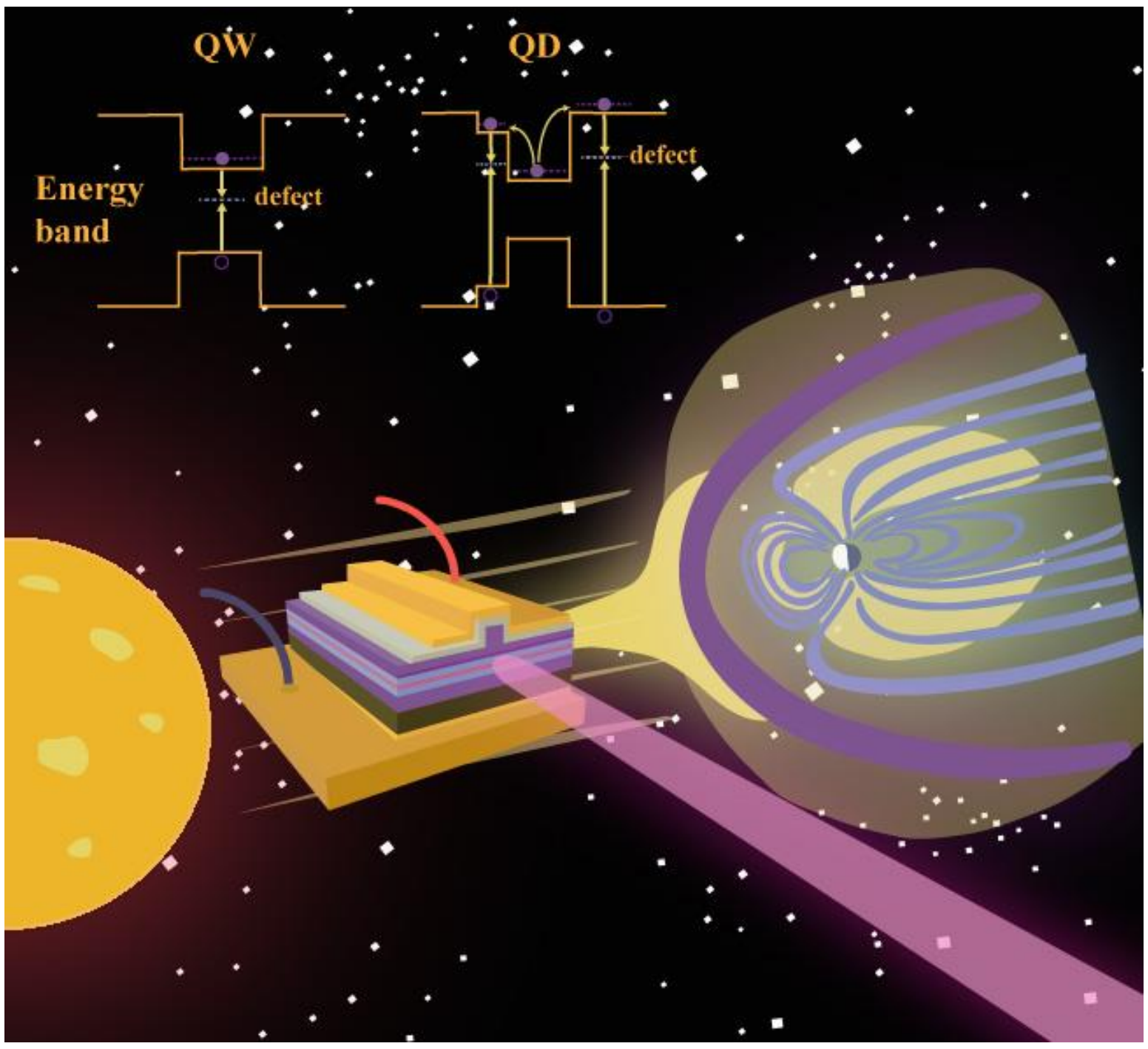
QW
QD
Energy
band
defect
defect